\documentclass[aps,showpacs,twocolumn,floatfix,amssymb,superscriptaddress,longbibliography]{revtex4-2}
\usepackage[toc,page]{appendix}
\usepackage[british]{babel}
\usepackage[utf8]{inputenc}
\usepackage{amsmath,amssymb}
\usepackage{graphicx}
\usepackage{ae}
\usepackage{esdiff}
\usepackage{braket}
\usepackage{bbm}
\usepackage{simplewick}
\usepackage{float}
\usepackage{braket}
\usepackage{simplewick}
\usepackage{color}
\usepackage{framed}
\usepackage[caption=false]{subfig}
\usepackage[pdftex,colorlinks=true,linkcolor=blue,citecolor=blue,urlcolor=black]{hyperref}
\usepackage{hyperref}
\usepackage{mathtools}
\usepackage{enumitem}
\usepackage{ragged2e}
\usepackage{floatflt}
\usepackage{import}
\usepackage{titlesec}
\usepackage{multirow}
\usepackage{etoolbox}
\usepackage{appendix}
\usepackage{dsfont}
\usepackage{comment}
\usepackage{xcolor}
\usepackage{blindtext}
\usepackage[makeroom]{cancel}
\usepackage{mathtools}
\usepackage{lineno}
\usepackage[normalem]{ulem}

\begin{document}

\newcommand{\abs}[1]{\mathopen|#1\mathclose|}

\definecolor{mgrey}{RGB}{63,63,63}
\definecolor{mred}{RGB}{235,97,51}
\newcommand{\mg}[1]{{\color{mgrey}{#1}}}
\newcommand{\mr}[1]{{\color{mred}{#1}}}
\def \k{{\mathbf{k}}}
\def \p{{\mathbf{p}}}
\def \q{{\mathbf{q}}}
\def \r {{\mathbf{r}}}
\def \Q {\mathbf{Q}}
\def \R {\mathbf{R}}
\def \x {\mathbf{x}}
\def \y {\mathbf{y}}
\def \a {\mathbf{a}}
\def \b {\mathbf{b}}
\def \d {\mathbf{d}}
\def \c {\mathbf{c}}
\def \bn {\boldsymbol{0}}
\def \intphi { \int \frac{d\phi_k}{2\pi} \frac{d\phi_p}{2\pi}}
\newcommand{\red}[1]{{\color{red}{#1}}}

\title{Polaron spectra and edge singularities for correlated flat bands} 

\author{Dimitri Pimenov}\email{dp589@cornell.edu}
\affiliation{Department of Physics, Cornell University, Ithaca, New York 14853, USA}

\begin{abstract}
Single- and two-particle spectra of a single immobile impurity immersed  in a fermionic bath can be computed exactly and are characterized by divergent power laws (edge singularities). Here, we present the leading lattice correction to this canonical problem, by embedding both impurity and bath fermions in bands with non-vanishing Bloch band geometry, with the impurity band being flat. By analyzing generic Feynman diagrams, we pinpoint how the band geometry reduces the effective interaction which enters the power laws; we find that for weak lattice effects or small Fermi momenta, the leading correction is proportional to the Fermi energy times the sum of the quantum metrics of the bands. When only the bath fermion geometry is important, the results can be extended to large Fermi momenta and strong lattice effects {and cross-validated by analysis of $S$-matrix eigenvalues}.  We numerically illustrate  our results on the Lieb lattice and draw connections to various spectroscopy experiments. 
\end{abstract}

\maketitle

\section{Introduction} 

A rare example of an analytically solvable quantum problem is the description of a single immobile impurity embedded in a Fermi sea \cite{PhysRev.163.612, PhysRev.178.1084,PhysRev.178.1072,RevModPhys.62.929}. Despite many-body appearances, this problem can be formulated in single-particle language \cite{PhysRev.178.1097, PhysRev.182.479,  combescot1971infrared, PhysRevLett.91.266602, PhysRevB.71.045326}: if the impurity is structureless, it acts as a time-dependent scattering potential for the Fermi sea electrons, inducing  phase shifts of  the single-particle orbitals. These lead to a vanishing overlap of the Fermi sea ground state  with and without impurity -- a phenomenon known as Anderson Orthogonality catastrophe \cite{PhysRevLett.18.1049}. The resulting impurity spectra feature divergent power laws (edge singularities), whose exponents can be expressed via the  phase shifts at the Fermi level.

Due to its exact solvability, the edge singularity setup can be a starting point to explore related problems which are undeniably many-body. One possible direction is to consider heavy but mobile impurities. The finite mass adds recoil,  cutting off the singularities in the problem \cite{gavoret1969optical, PhysRevB.4.4315, PhysRevB.35.7551, PhysRevLett.65.1048, PhysRevB.44.3821, nozieres1994effect, PhysRevLett.75.1988, PhysRevB.96.155310}. For equal masses, one arrives at the so-called ``Fermi-polaron" problem which has gained enormous traction in the context of ultracold gases and cavity semiconductor experiments in the last years \cite{schmidt2018universal, LevParpolreview, massignan2014polarons}. 

Another interesting route is the modification of edge singularities due to lattice effects. In previous studies of edge singularities for lattice models, only the Fermi sea fermions were subject to specific lattice or trap enviroment \cite{PhysRevB.76.115407, PhysRevLett.111.165303, PhysRevB.82.153410}. Here, we propose a new variant of the problem: both bath fermions and impurity are placed in bands with non-trivial band geometry. Such a situation can for instance arise if the impurity is a single heavy hole created by photoexcitation out of a flat band. We aim to determine the \textit{universal} leading modification of the edge singularities due to (weak) lattice effects, independent of the concrete lattice of choice. 

Non-trivial flat band systems form an ideal breeding ground for strong-correlation physics, since interactions dominate over kinetic effects. Typical examples are given by Landau Levels  and modern Moir\'e materials like TBG \cite{TBGCao1,Moirerev}. Another venue for generation of flat bands are optical lattices hosting ultracold atoms, where for instance the Lieb lattice has been realized  \cite{doi:10.1080/23746149.2018.1473052, doi:10.1126/sciadv.1500854,PhysRevLett.118.175301, taieLiebnew}.  While all these systems are most interesting at generic filling, they are typically inaccessible by controlled theory. The setup we are considering here is simpler: kinetic effects are still quenched, but the single-hole limit provides a way to get the interactions under control. 

Our theoretical starting point is a multiband model with two active bands, a flat one (termed $f$ band), which is initially filled, and a dispersive one (termed $c$ band) filled up to the chemical potential $\mu$. We assume a short-ranged interaction between $f$ and $c$ particles, which contains overlaps of Bloch functions (form factors) as a result of the lattice structure. Our goal is to compute single-hole correlation functions $A(\nu)  \sim \text{Im} \braket{f^\dagger  f}\!(\nu)$ and $\chi(\nu) \sim \text{Im} \braket{ c f^\dagger f c^\dagger}\!(\nu)$, which describe photoemission spectra (RF spectra in ultracold atom experiments) or interband absorption spectra, respectively.

In the standard edge singularity scenario, the spectra scale as $A(\nu) \sim \nu^{2\alpha^2 -1}, \chi(\nu) \sim \nu^{-2\alpha}$, where $\alpha$ is the dimensionless $c$-$f$ interaction. On the lattice, the form factors come into play. 
{For small $\alpha$, the modifications due to form factors can be  treated exactly if only the  $c$-band geometry is of importance, for general band fillings. This can be achieved either by evaluating Bloch overlaps for generic diagrams, or by computing  $S$-matrix eigenvalues in the Born approximation}.  If the $f$-band geometry is important as well, the single-particle character of the problem is lost in general: For instance,  a finite effective mass is generated for the $f$-hole. On the other hand, the problem can be controlled if the Fermi momentum $k_F$ is small, equivalent to a weak lattice effect: as we show by diagrammatic analysis, in this case the dominant effect is to reduce the interaction $\alpha$, while other effects, such as $f$-band mass generation, are subleading. The  interaction correction to $\alpha$ scales as $k_F^2 \text{tr} (g^f + g^c)$, where $g^c, g^f$ are the quantum metrics of the respective bands at the $c$-band minimum. If the photocreated hole has a momentum $|\Q| \gg k_F$, the $f$-band metric must be evaluated at this momentum. The reduced interaction can be interpreted in terms of the minimal real-space spread of the wavefunctions, which cannot be localized completely due to the Bloch band geometry: therefore, the effective scattering potential seen by the $c$ fermions acquires a finite range, reducing the phase shift at the Fermi level.
We check these results numerically for fermions moving in a weakly doped Lieb lattice, which contains a flat and dispersive band with a non-trivial metrics.

While Moir\'e materials can in principle host the lattice edge singularity physics, experimental observation might be challenging due to the insufficient energy resolution of spectral measurements to date. On the other hand, ultracold gas systems provide both a platform to realize topological flat bands and also come with a well-developed experimental toolbox to resolve polaronic spectra. In particular, we expect the discussed geometrical effects to be observable in RF spectroscopy \cite{PhysRevLett.99.170404,PhysRevA.102.023304}.

The remainder of this article is structured as follows: in Sec.\ \ref{setupsection}, we introduce the general Hamiltonian, and the Lieb lattice in Sec.\ \ref{Lieblatticesec}. In Sec.\ \ref{recapsec}, we recapitulate the standard continuum  edge singularities. In Sec.\ \ref{conly}, we introduce a band geometry for the $c$-band, study its impact on photoemission and absorption-type spectra, and introduce a perturbative formulation of lattice effects in terms of the quantum metric. In Sec.\ \ref{fullgeom}, we apply this perturbative treatment to the case where both $c$ and $f$ bands have a non-vanishing band geometry. In Sec.\ \ref{discsec} we discuss the experimental relevance and limitations of our results, and close in Sec.\ \ref{concsec} by providing a summary and  an outlook. Technical details are relegated to Appendices.

\section{Setup} 
\subsection{General Hamiltonian} 
\label{setupsection}
\begin{figure}
\centering
\includegraphics[width= 0.95\columnwidth]{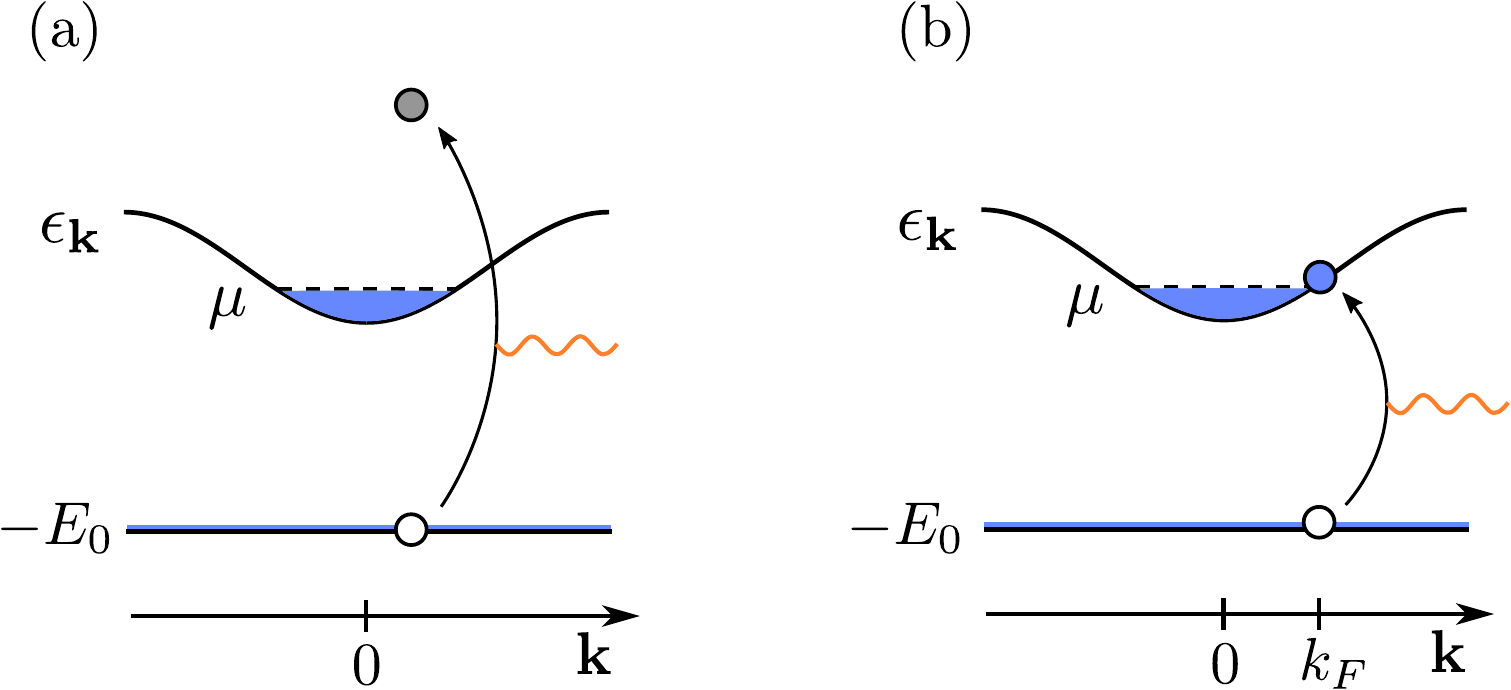}
\caption{Hole creation by photoexcitation. Filled (empty) circles indicate holes (electrons), the light pulse is indicated by a wiggly orange line.  (a) Photoemission, where the electron is ejected from the system (shown in gray) (b) Interband absorption, which creates a hole with momentum $\geq k_F$.   }
\label{bandfig}
\end{figure}

Consider a generic $d$-dimensional multi-band model at $T = 0$ with two \textit{active} bands: a flat valence band ($f$-band) and dispersive conduction band ($c$-band). In the band basis, the kinetic Hamiltonian reads: 
\begin{align} 
\label{Hkin}
H_\text{kin} = \sum_{\k} -E_0 f^\dagger_\k f_\k  +  \sum_\k \epsilon_\k c^\dagger_\k c_\k \ . 
\end{align} 
We use units such that $\hbar, e = 1$, and suppress spins -- they only lead to factors of two which we will reinstall where needed. Energies are measured from the bottom of the conduction band,  and we assume that the band gap $E_0$ is the largest scale in the problem, $E_0 \rightarrow \infty$.  As sketched in Fig.\ \ref{bandfig}, the dispersive band is occupied up to the chemical potential $\mu$ {(for results on Orthogonality Catastrophe for band insulators, see Ref.\ \cite{PhysRevB.94.125111})}. The flat band is filled as well, but these filled $f$ states are inert. We will study processes where a single  hole  is injected in the $f$ band. This can e.g.\ be achieved via photoexcitation in one of two ways: either, a high-energy light pulse is applied which ejects an $f$-particle from the system [photoemission, Fig.\ \ref{bandfig}(a)] or the $f$-particle is lifted into the $c$-band [interband absorption, Fig.\ \ref{bandfig}(b)] by irridating  light with an energy $\gtrsim (E_0+\mu)$.  

Now we include interactions. A general interaction term involving $c$,$f$ fermions can be written down as  
\begin{align} 
H_\text{int} &= \frac{1}{2} \int d^d r d^d r^\prime \Psi^\dagger(\r) \Psi^\dagger(\r^\prime) V(\r - \r^\prime) \Psi(\r^\prime) \Psi(\r)  \  , 
\end{align}
with field operators 
\begin{align} 
&\Psi(\r) = \frac{1}{\sqrt{\Omega}}\sum_{\k,n\in \{{f,c}\}} u_{n,\k} (\r) e^{i\k\cdot \r} a_{n,\k}\  .
\end{align} 
Here,  $ a_{f,\k} = f_\k, \ a_{c,\k} = c_\k$, $u_{n,\k}(\r)$ is the normalized cell-periodic Bloch function, and $\Omega$ is the system volume. In momentum space, the interaction becomes 
\begin{align} \notag
H_\text{int} = &\frac{1}{2\Omega} \sum_{\substack{n_1,n_2,n_3,n_4 \\ \k, \k^\prime, \q, {\mathbf{K}} }} V_{\q+{\mathbf{K}}} a^\dagger_{n_1,\k+\q} a^\dagger_{n_2,\k^\prime-\q} a_{n_3,\k^\prime } a_{n_4, \k} \\ \times & \braket{n_1,\k +\q| n_4,\k }_{\mathbf{K}} \braket{ n_2, \k^\prime  - \q |   n_3,  \k^\prime}_{-{\mathbf{K}}} , \label{fullinteraction} 
\end{align} 
where
\begin{align} 
&\braket{n_i,\k| n_j, \p}_{\mathbf{K}} = \\ &\frac{1}{\text{unit cell vol.\ } }  \int_\text{u.c.} d^d r \  \overline{{u_{n_i, \k}(\r)}} u_{n_j,\p}(\r) \exp(-i {\mathbf{K}}\cdot \r ) \ .  \notag
\end{align}

 $V_{\q+{\mathbf{K}}}$ is the interaction matrix element in momentum space. In Eq.\ \eqref{fullinteraction}, $\k, \k^\prime$ and $\q$ are restricted to the first Brillouin zone, while ${\mathbf{K}}$ is a reciprocal lattice vector. 
 
The band geometry is encoded in the Bloch factor overlaps in Eq.\ \eqref{fullinteraction}. To isolate their effects on the edge singularities, we will simplify the interaction as 
\begin{align} 
V_{\q +{\mathbf{K}}} \simeq V_0 \times \delta_{{\mathbf{K}},0} \ , 
\end{align}   
dropping the summation over ${\mathbf{K}}$. That is, we assume that the interaction in reciprocal space is essentially constant for the momentum transfers of interest (of order $k_F$), but decays quickly enough for momentum transfers on the order of a reciprocal lattice vector, which allows to neglect Umklapp processes. We assume that there is a clear separation between these scales, which applies in a weak doping limit. In real space, this implies that the interaction  is constant on the scale of a unit cell, but decays strongly on the much larger scale $1/k_F$. Note that this excludes interactions with an explicit sublattice structure. For a  screened Coulomb-like attraction between the $f$-hole and the $c$-electrons, $V_0 > 0$; {a discussion of edge singularities in the context of long-ranged Coulomb interactions can be found in Ref.\ \cite{PhysRevB.99.245122}}. To summarize, the most general interaction studied in this work reads 
\begin{align} \notag
H_\text{int} = &\frac{1}{2\Omega} \sum_{\substack{n_1,n_2,n_3,n_4 \\ \k, \k^\prime, \q }} V_0 a^\dagger_{n_1,\k} a^\dagger_{n_2,\k^\prime} a_{n_3,\k^\prime - \q} a_{n_4, \k + \q} \\ \times & \braket{n_1,\k| n_4,\k + \q} \braket{ n_2, \k^\prime | n_3,  \k^\prime - \q} \ .  \label{fullsimpleinteraction}
\end{align}

When a hole is created in a photoexcitation process [Fig.\ \ref{bandfig}(a)], the measured spectra are determined by the correlation functions involving $c,f$ operators. For the {photoemission} spectrum $A_\Q(\omega)$ (which can essentially be measured via RF spectroscopy in ultracold gas experiments \cite{PhysRevLett.99.170404, PhysRevA.102.023304}), the relevant correlation function is the  propagator: 
\begin{align} 
\label{singleholespec}
F_\Q(t) &= -i\braket{0|\hat{T} \left\{f_\Q(t) f^\dagger_\Q(0)\right\}|0}  \\ \notag
A_\Q(\omega) &= \text{Im} \left[F_\Q(-\omega) \right] \ , 
\end{align} 
where $\hat{T}$ is a time-ordering operator. In general, $\ket{0}$  is the interacting ground state; however, in the limit $E_0 \rightarrow \infty$ we are interested in, it is equivalent to the non-interacting ground state with a filled $f$-band. This implies that $F_\Q(t) \sim \theta(-t)$ is purely advanced when $E_0 \rightarrow \infty$. 
 
For the interband absorption spectrum $\chi(\omega)$ [Fig.\ \ref{bandfig}(b)], we need the interband current-current correlation function
\begin{align} \notag
&\Pi(t) = (- i) \times  \\& \sum_{\k_1, \k_2} \! J_{cf}^\eta(\k_2) J_{fc}^\eta(\k_1)   \braket{0 | \hat{T} \left\{ f_{\k_2}^\dagger(t) c_{\k_2}(t) c_{\k_1}^\dagger(0) f_{\k_1}(0)\right \} |0}  \notag \\ 
&\chi(\omega) = - \text{Im} \left[\Pi(\omega)\right] , 
\end{align}  
where the matrix elements of the interband operator can be expressed as \cite{tai2023quantum} 
\begin{align} \label{interbandcurr}
J_{cf}^\eta(\k) &= (E_0 + \epsilon_\k) \braket{c, \k| \partial_\eta | f,\k} , \quad \partial_\eta \equiv \partial_{\k_\eta} 
\\ &\simeq   E_0 \braket{c, \k| \partial_\eta | f,\k} \notag , 
\end{align} 
and the last approximation holds for large $E_0$. 

As pointed out in Ref.\  \cite{tai2023quantum}, care needs to be taken when evaluating optical response for a set of active bands, since interband current matrix elements scale with the band gap, see Eq.\ \eqref{interbandcurr}, and  higher ``passive" bands may therefore contribute as well. In our case, processes involving active bands  lead to logarithmic singularities when the external frequency is close to a specific threshold energy. For off-resonant higher-band contributions, such singularities should not appear, and we therefore neglect them in the following.

\subsection{Exemplary tight binding model: Lieb Lattice} 
\label{Lieblatticesec}
\begin{figure}
\centering
\includegraphics[width= \columnwidth]{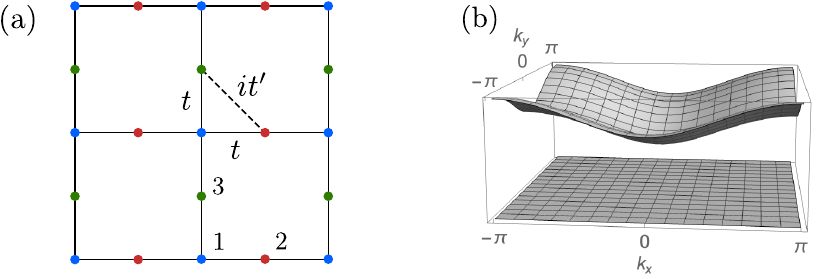}
\caption{Lieb lattice. (a) Lattice structure; sublattices are indicated by coloring. (b) Band structure for $t = 1, t^\prime = 0.6$ (bottom band is not  shown). Momenta are shifted by $(\pi,\pi)$.}
\label{Liebfig1}
\end{figure}

To illustrate our results on a concrete tight binding model, we will consider the two-dimensional Lieb lattice (see, e.g., Ref.\  \cite{PhysRevB.86.195129}), which has already been realized with optical lattices \cite{doi:10.1080/23746149.2018.1473052, doi:10.1126/sciadv.1500854,PhysRevLett.118.175301, taieLiebnew}. 
 The kinetic part of the Hamiltonian is given by 
\begin{align} 
& H_\text{Lieb} =   -  t\sum_{\R, s = \pm} a^\dagger_{2, \R + s \hat{\x}} a_{1, \R} + a^\dagger_{3, \R + s \hat{\y}} a_{1,\R}    \\ & + it^\prime \sum_{\R, s = \pm} a^\dagger_{3,\R + s\hat{\y}} a_{2,\R + s\hat{\x}} + a^\dagger_{2,\R - s\hat{\x}} a_{3,\R + s\hat{\y}} + \text{h.c.} ,  \notag 
\end{align}  
where $a_1, a_2, a_3$ denote operators on sublattices $1,2,3$ as indicated in Fig.\ \ref{Liebfig1}(a). This lattice is characterized by an exactly flat central band, ($f$ band), and two dispersive bands, with the bandgap equal to $t^\prime$ [Fig.\ \ref{Liebfig1}(b)].  We consider a doped upper band, calling it the $c$-band. The minimum of the $c$-band is located at the $M$-point, which we will choose as momentum space origin for convenience.  For $t^\prime > 0$, the Lieb lattice bands are characterized by Chern numbers $(-1,0,1)$ \cite{PhysRevB.86.195129}. While the $f$ band has a vanishing Chern number, its Bloch functions $u_f(\k)$ are strongly varying.  

\section{Recap: Edge singularities for trivial bands} 
\label{recapsec}

To set the stage for evaluation of the spectra $A_\Q(\omega), \chi(\omega)$ in the lattice case, we recapitulate the standard continuum solution of the edge singularity problem \cite{PhysRev.163.612, PhysRev.178.1084,PhysRev.178.1072,RevModPhys.62.929,PhysRev.178.1097, combescot1971infrared,schmidt2018universal}. It assumes that the hole is  featureless and the momentum-dependence of hole operators can be erased, $f_\k \rightarrow f$. For the conduction band, the form factors are usually suppressed, and the interaction $H_\text{int}$ is approximated as 
\begin{align} 
H_\text{int}^{(0)} = {-}&\frac{1}{\Omega} \sum_{\k,\q} V_0 c^{\dagger}_{\k + \q} c_\k f {f^\dagger \label{standardint}} \ ,  
\end{align}
{where we have permuted the $f$-operators to ensure that the interactions are turned on in the presence of  holes. \footnote{This permutation should be performed before erasing momentum labels for the $f$-fermions,  and only leads to a constant energy shift for the $c$-fermions}}. 
Only $c$-$f$ interactions are retained in Eq.\ \eqref{standardint}. In particular, interaction terms involving {$c$}-fermions only are neglected, assuming that they lead to a renormalization of the {$c$}-band that can be absorbed in the bare Hamiltonian. Furthermore, one typically assumes rotational invariance. 

{Given that the operator $f f^\dagger$ can only take the values $0$ when the hole is absent and $1$ if it is present, the Hamiltonian is effectively quadratic, and the 
evaluation of spectra can be reduced to solving a time-dependent scattering problem. A number of exact approaches have been developed to this end, which rely on the solution of singular integral equations \cite{PhysRev.178.1097,combescot1971infrared}, bosonization \cite{PhysRev.182.479} or the solution of  Riemann-Hilbert boundary value problems \cite{PhysRevLett.91.266602, PhysRevB.71.045326}. The resulting spectra feature divergent power laws (``edge singularities"), whose exponents are determined by the phase shifts $\delta$ of conduction electrons on the Fermi surface. } For the momentum-independent ($s$-wave) interaction of Eq.\ \eqref{standardint}, one obtains: 
{
\begin{align}
\label{strongcouplingps}
&A(\nu) \sim \nu^{2(\delta/\pi)^2  -1}\  , &&\nu \equiv \omega - E_0    \\  \notag
&\chi(\nu) \sim \nu^{-2\delta/\pi + 2(\delta/\pi)^2} \ ,  &&\nu \equiv \omega - E_0 - \mu \ , 
\end{align} 
}
where $\nu$ is the energy measured from the respective threshold, neglecting perturbative threshold shifts, and non-singular dependencies on $\nu$. The factor $2$ {in the exponents which multiplies the $\delta^2$ terms (but not the prefactor of the linear term)} comes from spin. Frequencies are measured in units of a UV cutoff $\Lambda$ of order $\mu$. The results \eqref{strongcouplingps} are is asymptotically exact as $\nu \rightarrow 0$.

For  momentum-dependent scattering potentials, the scattering phase shifts can be defined via the eigenvalues $\sim \exp(i2\delta_j)$ of a (time-independent) $S$-matrix at the Fermi energy \cite{PhysRevLett.91.266602, PhysRevB.71.045326}. The spectra become 
\begin{align}
&A(\nu) \sim \nu^{2\sum_j  (\delta_j/\pi)^2 -1} \label{partialpsA} \\
&\chi(\nu) \sim \sum_{j} \nu^{-2\delta_j /\pi + 2 \sum_{j^\prime} (\delta_{j^\prime}/\pi)^2}  \label{partialpschi} \ , 
\end{align}

where in the last equation we suppressed the prefactors of the power laws in the various channels. For a spherically symmetry potential, the channel indices correspond to angular momenta \cite{PhysRev.178.1097}.

While the non-perturbative solutions are elegant, they are difficult to generalize to many-body variants of the impurity problem. Instead, one can take the diagrammatic solution (which was found first historically \cite{PhysRev.178.1072}) as a starting point. The Feynman diagrams are organized in powers of an effective dimensionless coupling constant 
\begin{align} 
\alpha \equiv \rho V_0 \ , 
\end{align} 
where $\rho$ is the density of states at the Fermi level. For $\alpha \ll 1$, the spectra obtained via summation of diagrams read, to the leading order in $\alpha$: 
\begin{align}
&A(\nu) \sim \nu^{2\alpha^2 -1 }\theta(\nu)   \label{Aleading} \\ 
&\chi(\nu) \sim \nu^{-2\alpha} \theta(\nu) \label{chileading} \ , 
\end{align} 
where $\delta \sim \pi \alpha$ for $\alpha \ll 1$, such that Eq.\ \eqref{strongcouplingps} agrees with the weak-coupling result in this limit (the exact functional form of $\delta(\alpha)$ depends on dimensionality and UV regularization). 

\begin{figure}
\centering
\includegraphics[width= \columnwidth]{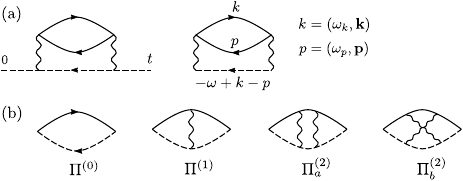}
\caption{{Feynman diagrams  that determine the photoemission $A$ and the interband absorption $\chi$. Dashed and full lines correspond to $f$ (c) particles, respectively, and wavy lines to the interaction. }(a) Left: Leading contribution to the hole propagator $C_2(t)$, which is exponentiated in the linked cluster approach. Right: Self-energy part in frequency domain. (b) Leading parquet diagrams that determine $\Pi(\omega)$ up to second order in the interaction.}
\label{Diagfigfirst}
\end{figure}

{The weak-coupling results} can be derived by summing the leading logarithmically divergent (parquet) diagrams. An essential fact is that the asymptotic behavior as $\nu \rightarrow 0$ is dominated by scattering of $c$-fermions  close to the Fermi surface. For $A\sim \nu^{\alpha^2 -1}$, the summation of diagrams is most easily achieved by applying the ``linked cluster" theorem \cite{mahan2000many}, which states that 
\begin{align} 
\label{exponentiation} 
F(t) \sim \exp\left[\sum_n C_n(t)\right] \  . 
\end{align} 
$C_n(t)$ can be related to time-dependent Feynman diagrams. The leading behavior, Eq.\ \eqref{Aleading}, derives from the the second-order diagram shown in Fig.\ \ref{Diagfigfirst}(a), which is evaluated as $C_2(t) =  - \alpha^2 \log(\Lambda |t|)$ as $|t| \rightarrow \infty$ \footnote{ up to factors $\theta(-t) e^{iE_0 t}$ which we suppress here for brevity}. The evaluation works with logarithmic accuracy, neglecting $O(1)$ terms when compared to large logarithms. Fourier transform of $\exp[C_2(t)]$ leads to the result in Eq.\ \eqref{Aleading}. 

For the derivation of $\chi \sim \nu^{-2\alpha}$, a short-cut as in Eq.\ \eqref{exponentiation} is not available, and the result {is} computed by solving a set of coupled Bethe-Salpeter equations \cite{PhysRev.178.1072}. The first few relevant diagrams in frequency space are shown in Fig.\ \ref{Diagfigfirst}(b); they reproduce a perturbative expansion \cite{PhysRev.163.612}: 
\begin{align} \notag
\Pi(\nu) &= \sum_\eta  \frac{|J^\eta(k_F)^2| \rho}{\alpha}\left( \alpha L - \alpha^2 L^2 + \frac{2}{3} \alpha^3 L^3 + \hdots \right)  \\&\simeq \sum_\eta \frac{|J^\eta(k_F)^2| \rho}{\alpha} \left[ 1 -  \exp(-2 \alpha L)\right], 
\label{Pinupert}
\end{align} 
where $L = \log(|\nu|/\Lambda)$; $\chi(\nu)$ can be derived by restoring the correct imaginary part of the logarithms via Kramers-Kronig relations. To the leading order in $\alpha$, self-energy  diagrams or vertex corrections do not contribute to $\chi$.

In the preceding discussion, we have implicitly assumed the absence of bound states. These can lead to additional low-energy singularities in the spectra, with associated power laws that have a form similar to Eq.\ \eqref{strongcouplingps} \cite{combescot1971infrared,schmidt2018universal}; in this case, the phase shifts are close to $\pi$. To recover these results in a diagrammatic calculation, the interaction lines in Fig.\ \ref{Diagfigfirst} need to be replaced by ladders ($T$-matrices), see, e.g., Ref.\ \cite{PhysRevB.98.220302}. To keep the computations manageable, we will neglect the effect of bound states, which is justified for sufficiently small binding energies. 

\section{$c$-band geometry only} 
\label{conly}

In the following, our goal is to relax the assumptions that lead from $H_\text{int}$, Eq.\ \eqref{fullsimpleinteraction} to $H_\text{int}^{(0)}$, Eq.\ \eqref{standardint}, step by step. To begin with, we keep the hole structureless, but allow for a non-trivial $c$-band geometry, using an interaction of the form 
\begin{align} 
H_\text{int}^{(1)} = {-}&\frac{1}{\Omega} \sum_{\k,\q} V_0 c^{\dagger}_{\k + \q} c_\k {f f^\dagger}  \label{Hint1}   \braket{c,\k+ \q| c,\k} \ . 
\end{align}
Such an interaction can be appropriate if the $c$ and $f$ electrons are in fact different particle species, as is the case in typical polaron-type ultracold gas experiments. 

The interaction term \eqref{Hint1} also describes a scattering potential for the $c$ fermions. { When the $f$-hole is present, $f f^\dagger = 1$,  we can rewrite Eq.\ \eqref{Hint1} in single-particle notation as 
\begin{align} 
\label{singlepartH1}
H_\text{int}^{(1)} &= -\frac{1}{\Omega} \sum_{\k,\q} V_0\braket{c,\k + \q|c, \k}  \ket{\Psi_{c,\k+\q}}\! \bra{\Psi_{c,\k}}    \\
& \notag = -\frac{1}{\Omega} \sum_{\k,\q} V_0 e^{i(\k+\q)\hat{\r}} P_{\k + \q}^c P^c_{\k} e^{-i \k \hat{\r} } \ . 
\end{align} 

Here, $\Psi_{c, \k} = \exp(i\k \hat{\mathbf{r}}) \ket{\k}$ is the full energy eigenstate (i.e, not just the cell-periodic part), $\hat{\r}$ is the position operator, and $P_\k^c = \ket{\k}\! \bra{\k}$ is the projector on the $c$-band Bloch  function with momentum $\k$. The formulation \eqref{singlepartH1} has the advantage of being manifestly gauge invariant, and is a convenient starting point for an analysis of the scattering problem.

Here, we will instead follow the Feynman diagram approach, since it can readily be generalized to the case of a non-trivial $f$-band as discussed in the later Sec.\ \ref{fullgeom}. In Appendix \ref{Sapp} we show how to recover the same weak-coupling results via $S$-matrix analysis using Eq.\ \eqref{singlepartH1}. 
 }


\subsection{Single-hole spectrum $A$}

To analyze the single-hole (photoemission)  spectrum $A(\nu) = \text{Im}[F(-\nu)]$, consider the self-energy part of the second order diagram  for the hole propagator $F$ evaluated in the \textit{imaginary} frequency domain [Fig.\ \ref{Diagfigfirst}(a)]: 
\begin{align} 
\notag
&\Sigma(-\omega) = -V_0^2 \int \frac{d\omega_k}{2\pi} \frac{d\omega_p}{2\pi} \frac{d\k}{(2\pi)^d} \frac{d\p}{(2\pi)^d}  \abs{\braket{\k|  \p }}^2 \\  &G_c(\omega_k, \k) G_c (\omega_p, \p) F^{(0)} (-\omega + \omega_k - \omega_p)  \ . \label{Sigmafirst}
\end{align} 
where $G_c(\omega, \k) = (i\omega - (\epsilon_k - \mu))^{-1}, F^{(0)}(\omega) = (i\omega + E_0)^{-1}$. Evaluating the frequency integrals, one obtains 
\begin{align} 
&\Sigma(-\omega) = \\ &V_0^2 \int_{\substack{\epsilon_\k > \mu \\ \epsilon_\p < \mu}} \frac{d\k}{(2\pi)^d} \frac{d\p}{(2\pi)^d} \frac{1}{-i\omega + E_0 + \epsilon_\k - \epsilon_\p} \abs{\braket{\k|  \p }}^2 \notag \  . 
\end{align} 
After analytical continuation $i\omega \rightarrow \omega + i0^+$, we can e.g. evaluate the imaginary part as 
\begin{align} 
 \notag   \text{Im}\left[\Sigma(-\nu) \right] &\simeq \pi \alpha^2 \int_{0}^\mu d\epsilon_\p \int_\mu^\Lambda d\epsilon_\k \ \delta(\nu -(\epsilon_\k - \epsilon_\p))  \cdot I_2 \\   & = {\pi \alpha^2 \theta(\nu)  \nu }  \cdot I_2   \quad \text{as} \  \nu \rightarrow 0  . \label{leadingfreq}
 \end{align} 
 Here, $I_2$ corresponds to the gauge-invariant squared overlap of $c$-Bloch functions averaged over the Fermi surface: 
 \begin{align} 
 \label{I2}
I_2 \equiv \int_{\k,\p}  \abs{\braket{\k |  \p}}^2  , \quad \int_\k \equiv \frac{1}{\rho} \int \frac{d\k}{(2\pi)^d} \delta( \mu - \epsilon_\k) \ . 
\end{align}
The restriction to the Fermi surface alone is an approximation for general $\nu$, but it becomes exact as $\nu \searrow 0$ in Eq.\ \eqref{leadingfreq}. By the Kramers-Kronig relations, $\text{Re}\left[\Sigma(-\nu)\right] - \text{Re}[\Sigma(0)] \sim \nu \log(|\nu|)$, showing the emergence of logarithmic factors. Note that internal momenta that contribute to $\text{Re}[\Sigma(0)] \sim \alpha^2 \mu \log(\mu)$ do not have to be restricted to the proximity of the Fermi surface, as can be seen from direct computation or Fumi's theorem \cite{mahan2000many}; however, $\text{Re}[\Sigma(0)]$ is only essential for determination of  threshold energies, but not for the detailed form of the spectra. 

{In terms of the projector $P_\k^c = \ket{\k}\!\bra{\k}$},  we can rewrite 
\begin{align} 
\notag
I_2 &= \int_{\k,\p} \text{Tr} \left\{ P_{\k}^c P_{\p}^c \right\} = \text{Tr} \{ \left(P^c\right)^2 \} < 1 \\   \label{FSprojector}
P^c &\equiv \int_{\k} P^c_\k  \ , 
\end{align} 
where the trace acts in band ({or orbital}) space, and $P^c$ is the projector averaged over the Fermi surface. We therefore see that at the level of the second order diagram, the only relevant change we need to perform is $\alpha^2 \rightarrow \alpha^2  I_2$. 

The derivation of the linked cluster theorem for $F(t)$, Eq.\ \eqref{exponentiation}, relies only the $f$-hole being structureless, which equally applies to the interaction \eqref{Hint1}\cite{mahan2000many}. In evaluating the relevant expression $C_2(t)$, one then obtains the same result as in the infinite mass case, but again with the replacement $\alpha^2 \rightarrow \alpha^2 I_2$. Therefore, to the leading order in $\alpha$, but treating the $f$-band geometry exactly, the spectrum becomes: 
\begin{align} 
\label{Anucgeom}
A(\nu) \sim \nu^{2\alpha^2  \text{Tr} \{ \left(P^c\right)^2 \} -1} \ . 
\end{align}

The  interaction is suppressed by the Bloch-overlaps which penalize large momentum transfers. This suppression is enhanced for larger doping $k_F \nearrow$, since the overlap integral  $\text{Tr} \{ \left(P^c\right)^2 \}$ probes larger momenta. Therefore, the photoemission spectrum gets more singular. Two representative plots for the Lieb lattice are shown in Fig.\ \ref{Liebnum1}(a). For $\alpha^2  \text{Tr} \{ \left(P^c\right)^2 \} \rightarrow 0$, one recovers the non-interacting form $A(\nu) = \delta(\nu)$ when restoring normalizing factors.

\subsection{Particle-hole spectrum $\chi$}

\begin{figure}
\centering
\includegraphics[width= \columnwidth]{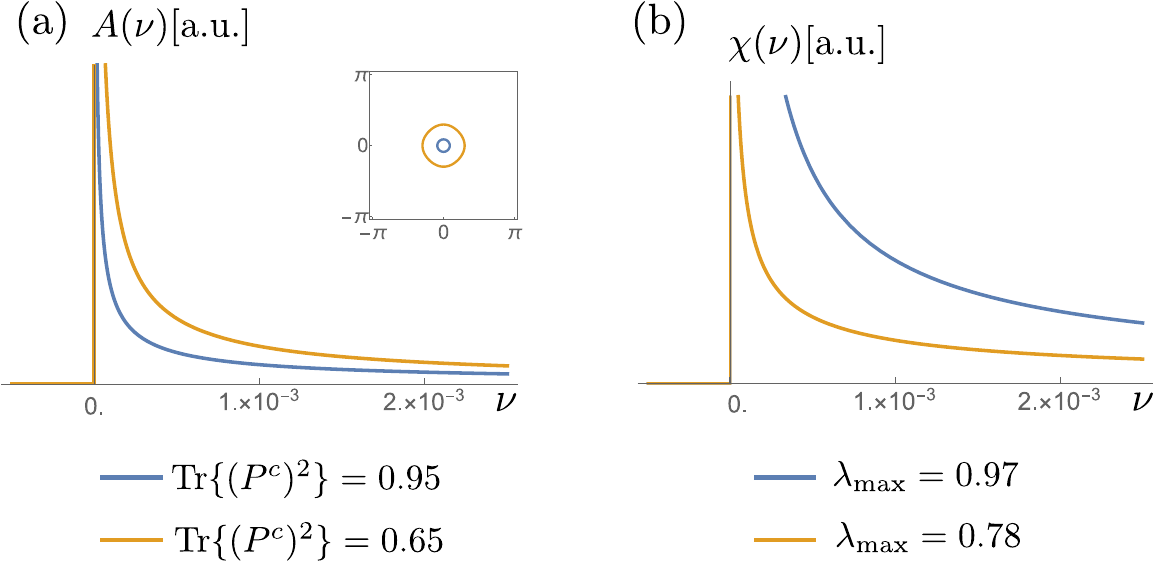}
\caption{ Edge singularities on the Lieb lattice, including $c$-band geometry. Used parameters: ${\alpha = 0.4}, t^\prime = 0.6$. (a) Single-hole spectrum $A(\nu)$, Eq.\ \eqref{Anucgeom} in arbitrary units; frequencies are measured in units of the UV cutoff. The inset shows the respective Fermi Level.  (b) Absorption spectrum $\chi(\nu)$, Eq.\ \eqref{Anunorm}. The plot shows the most singular contribution with the maximal Eigenvalue $\lambda_\text{max}$. }
\label{Liebnum1}
\end{figure}

For the particle-hole (intraband absorption) spectrum, we consider the correlation function 
\begin{align} \notag
\Pi(t) &= (- i)  \sum_{\k_1, \k_2, \eta} \! E_0^2 \braket{\k_1 | \hat{O}^\eta(\k_1,\k_2)  |  \k_2}  \times \\ &\braket{0 | \hat{T} \left\{ f^\dagger(t) c_{\k_2}(t) c_{\k_1}^\dagger(0) f(0) \right \} |0} \notag  \\  
\hat{O}^\eta(\k_1,\k_2) &=  - \partial_\eta \ket{ f, \k_1} \!   \left( {\partial_\eta} \bra{ f, \k_2} \right) \
\end{align} 
To maintain the assumption of a structureless hole, we assume that the momentum-dependence of the operator $\hat{O}^\eta$ is negligible. One can now evaluate the $\Pi(\nu)$ in similar manner as $\Sigma$ described in the previous section, for instance by computing the leading logarithmic diagrams of Fig.\ \ref{Diagfigfirst}(b). In this computation, an  independent momentum variable can be chosen for each $c$-fermion line. As a result, the low-order diagrams incur the following additional gauge-invariant factors due to the Bloch overlaps: 
\begin{align} 
&\Pi^{(0)}\!:  &&\sum_\eta \int_{\k} \braket{ \k | \hat{O}^\eta | \k } = \sum_\eta \text{Tr}\{\hat{O}^\eta P^c \}  \\ \notag
&\Pi^{(1)}\!:   &&\sum_\eta \int_{\k,\p} \braket{ \k | \hat{O}^\eta |   \p} \braket{ \p |   \k} =  \sum_\eta \text{Tr}\{\hat{O}^\eta (P^c)^2 \} \\ \notag
&\Pi^{(2)}_{a,b}\!: && \sum_\eta \text{Tr}\{\hat{O}^\eta (P^c)^3 \}  \ , 
\end{align} 
with $P^c$ the Fermi-surface averaged projector introduced in Eq.\ \eqref{FSprojector}. 
This structure continues for all diagrams without additional fermion loops; all leading parquet diagrams are of this form \cite{PhysRev.163.612, PhysRev.178.1072}. Therefore, we can rephrase the perturbative expansion for $\Pi(\nu)$, Eq.\ \eqref{Pinupert} as 
\begin{align}  
&\Pi(\nu) =  \\ &\sum_\eta \text{Tr} \left[ \frac{\hat{O}^\eta \rho}{\alpha} \left( \alpha L P^c - (\alpha L)^2 (P^c)^2 + \frac{2}{3} (\alpha L )^3 (P^c)^3  \hdots\right)  \right] . \notag
\end{align} 
Summing this series, we obtain 
\begin{align} 
\label{Pinucgen}
\Pi(\nu) = \sum_\eta \text{Tr} \left[ \frac{\hat{O}^\eta \rho}{\alpha} \left( \mathbbm{1} - \exp(-2\alpha L(\nu) P^c \right)\right] \ , 
\end{align} 
where $\mathbbm{1}$ is the identity in band space and the exponential is a matrix-exponential. Diagonalizing $P^c$, we obtain for $\chi(\nu) = \text{Im} [ \Pi(\nu) ]$:  
\begin{align} 
\label{chinucgeom}
\chi(\nu) \sim \sum_{i} \nu^{-2\alpha \lambda_i}, 
\end{align} 
where $\lambda_i$ are the Eigenvalues of $P^c$, and we suppressed the prefactors of the the power laws which correspond to the diagonal elements  of $\hat{O}^\eta$ rotated into the $P^c$ Eigenbasis. 
Note the analogy between Eqs.\ $\eqref{partialpsA}, \eqref{partialpschi}$ and Eqs.\ \eqref{Anucgeom} and \eqref{chinucgeom}: $\chi$ is a sum of power laws, while $A$ is a single power law, with the sum in the exponent.
{In Appendix \ref{Sapp}, we re-derive these results by evaluating the $S$-matrix in the Born approximation. }

 {When $k_F$ is increased, the particle-hole spectrum $\chi$ becomes} \textit{less} singular because the maximal Eigenvalue  $\lambda_\text{max}$ in Eq.\ \eqref{chinucgeom} is reduced, as we analyze in detail in the next section. This behavior is expected, since $\chi$ simply becomes a step-function in the limit $\nu \rightarrow 0$ for a constant density of states. In Fig.\ \ref{Liebnum1}(b), we show $\chi(\nu)$ for two values of the chemical potential $\mu$, pinpointing this behavior.

\subsection{Perturbative expansion for small band geometry} 
For a simpler interpretation of the results, which is also generalizable to the case of non-trivial $f$-bands,  it is useful to consider the limit of ``weak" $c$-band geometry: we assume that the momentum-variation of Bloch functions $\ket{\k}$ is small. This can always be justified for weak doping $\mu$, when $k_F$ is much smaller than a reciprocal lattice vector which sets the typical scale for the variation of $\ket{\k}$. One can think of this limit as the leading lattice correction to the continuum limit. A related perturbative expansion in the context of excitons can for instance be found in Ref.\ \cite{PhysRevLett.115.166802}. 

Under this assumption, we can expand $\ket{\k}$ around $\k = \bn$, where $\bn$ denotes the momentum where the $c$-band has its minimum; however, with the same accuracy, any other momentum within the Fermi volume can be chosen as point of expansion. Up to second order in $\k$, we have   
\begin{align}
\ket{\k} \simeq \ket{\bn} + k_i \ket{i} + \frac{1}{2} k_i k_j \ket{ ij} , 
\end{align} 
with the notation $\ket{i} \equiv \partial_{k_i} \ket{\k}\big|_{\k = \bn}$. Applying this expansion to the squared overlap $|\!\braket{\k | \p}\!|^2$ which appears in $I_2 = \text{Tr} \{ (P^c)^2 \}$, Eq.\ \eqref{FSprojector}, we obtain 
\begin{align} \notag
|\!\braket{\k | \p}\!|^2 &\simeq 1 - g^c_{ij} (\k - \p)_i (\k - \p)_j \\ & \equiv 1 -  || \k - \p ||^2_c  \ . \label{cpexp}
\end{align}
Here,  $g_{ij}$ is the quantum (Fubini-Study) metric of the $c$ band at the point $\k = \bn$,  defined by 
\begin{align} 
\label{gdef}
&g_{ij}^c = \frac{1}{2} \left( \braket{i  | j} + \braket{j  | i} \right) + \braket{\bn | i}\braket{\bn| j} . 
\end{align} 
To get from Eq.\ \eqref{cpexp} to \eqref{gdef}, we used  the identities  $\braket{\bn |i} + \braket{ i | \bn} = 0$ and  $\frac{1}{2} \left(\braket{i | j} +  \braket{j | i} \right) = \text{Re} \braket{i | j} =  - \text{Re} \braket{ ij  | \bn}$.  

The quantum metric is the real part of the $c$-$c$ component of the quantum geometric tensor \cite{ PhysRevB.56.12847, cheng2010quantum, PhysRevB.87.245103, PhysRevB.90.165139, PhysRevB.104.045103}, while the imaginary part corresponds to the Berry curvature $B$. $g^c$ is a gauge-invariant positive-semidefinite measure of the distance of the Bloch states in Hilbert space. The Brillouin-zone average of $\text{tr}g^c(\k)$ can be related to the minimal real-space spread $\braket{\r^2} - \braket{\r}^2$ of Wannier functions \cite{PhysRevB.56.12847}. While $g^c$ can be non-vanishing for a topologically trivial band, for a band with non-zero Berry curvature $B$ it is bounded from below: $\text{tr} g^c(\k) \geq |B(\k)|$ for any $\k$ \cite{PhysRevB.90.165139}; lower bounds related to symmetries can also be derived as well \cite{PhysRevLett.128.087002}.  For the Lieb lattice, $\text{tr} g^c(\k)$ has a broad maximum at the $M$ point for $t^\prime \lesssim 1$ (for $t^\prime \rightarrow 0$, the metric is sharply peaked at the $M$ point), see Fig.\ \ref{Liebnum3}(a). Various theoretical proposals \cite{ PhysRevB.87.245103, PhysRevB.88.064304, PhysRevB.97.201117,  PhysRevLett.124.197002, PhysRevB.104.195133} and successful experiments \cite{PhysRevLett.122.210401, qmdiamond} to determine the quantum metric have been put forward.

\begin{figure}
\centering
\includegraphics[width= \columnwidth]{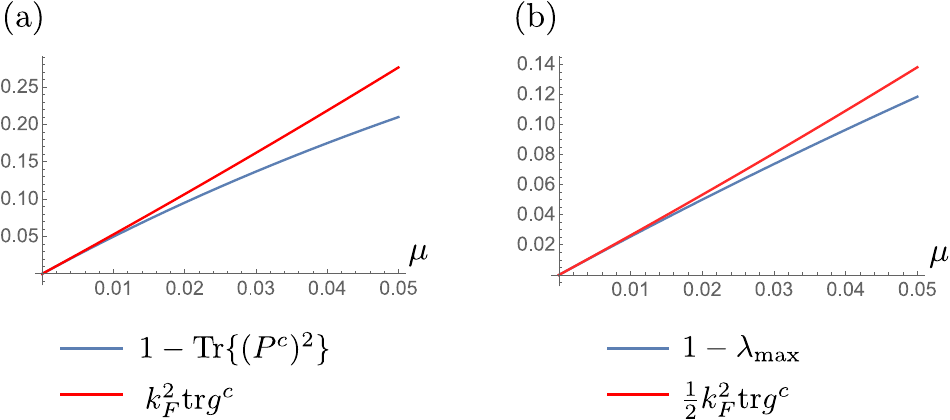}
\caption{(a) Evolution of lattice-corrections to the edge singularity exponents for the Lieblattice vs.\ chemical potential $\mu$, for $t^\prime = 0.6$. (a) Corrections relevant for $A(\nu)$. (b) Corrections relevant for $\chi(\nu)$.  }
\label{Liebnum2}
\end{figure}

With Eq.\ \eqref{cpexp} at hand, we can rewrite our result for $A(\nu)$,  Eq.\ \eqref{Anucgeom},  as 
\begin{align} 
\label{Anunorm}
A(\nu) \sim \nu^{2\alpha^2 (1 - \int_{\k,\p} || \k - \p ||_c^2) - 1} \ . 
\end{align}  
The edge-singularity is preserved, but the effective interaction is reduced by the Fermi-surface averaged Hilbert space distance of the scattered $c$-fermions. The expansion in Eq.\ \eqref{Anunorm} is controlled to first order in 
\begin{align} 
x_c\equiv k_F^2 \text{tr}g^c \overset{!}{\ll} 1\   , 
\end{align} where $k_F$ is a Fermi-surface averaged Fermi momentum \footnote{Note that $|g^c_{ij}| < \text{tr} g^c$ for a positive semidefinite matrix $g^c$.}. This parameter quantifies the leading lattice effect. 
For the Lieb lattice, the Fermi surface becomes nearly circular for small $k_F$, and the perturbative correction  simply reads
\begin{align} 
\int_{\k,\p} || \k - \p ||_c^2 \ {\rightarrow}  \ x_c  \quad \text{for} \ k_F \rightarrow 0, 
\end{align} 
as verified in Fig.\ \ref{Liebnum2}(a). This implies that, in principle, measuring the doping dependence of the edge-singularity exponent gives access to $\text{tr} g^c$ provided that the $k_F$-dependence of the coupling constant $\alpha$ is accounted for.

It is interesting to compare the result in Eq.\ \eqref{Anunorm} with the effect of a finite $f$-mass on $A(\nu)$ \cite{PhysRevLett.75.1988, nozieres1994effect, PhysRevB.96.155310}. Similar to Eq.\ \eqref{Anunorm}, the finite hole mass penalizes large momentum transfers which come at a high kinetic energy cost. This effect can be captured by evaluating an  phase space factor similar to $I_2$. However, in the case of a finite hole mass, this  factor vanishes as $\nu \rightarrow 0$ with a power dependent on dimensionality. This leads to a more drastic reorganization of the spectrum $A(\nu)$ compared to Eq.\ \eqref{Anunorm}, with a partial reemergence of the non-interacting delta function.

Similar to $A(\nu)$, an expansion in terms of the metric can also be applied to $\chi(\nu) \sim \sum_i \nu^{-2 \alpha \lambda_i}$, Eq.\ \eqref{chinucgeom}. We focus on the most singular contribution, i.e., the largest Eigenvalue $\lambda_{\text{max}} $. When $k_F \rightarrow 0$, $P^c$ becomes a  projector, and its largest Eigenvalue is $1$. For small $k_F$, second order perturbation theory (see App.\ \ref{eigenvalueshiftapp}) leads to 
\begin{align} 
\label{lambdamaxresult}
\lambda_\text{max} \simeq 1 - \int_{\k,\p} \frac{|| \k - \p ||_c^2}{2} \ , 
\end{align} 
which we check in Fig.\ \ref{Liebnum2}(b). Thus
\begin{align} 
\label{nukfcorr}
\chi(\nu) \sim \nu^{-2\alpha \left(1-  \int_{\k,\p} \frac{|| \k - \p ||_c^2}{2}\right)} \ . 
\end{align} 
Like in the photoemission case, in the absorption case the effective coupling constant is reduced. Note that, to the leading order in $x_c$, the same effective coupling
\begin{align}
\label{alphaeff}
\alpha_\text{eff} = \alpha\left(1 - \int_{\k,\p} \frac{|| \k - \p ||_c^2}{2}\right)
 \end{align} 
 appears in both Eqs.\ \eqref{Anunorm}, \eqref{nukfcorr}. Interestingly, this result agrees with a Fermi-surface average of the \textit{modulus} of the overlap term  in $H_\text{int}^{(1)}$ at order $O(x_c)$, with both incoming and outgoing momenta on the Fermi surface.

With the accuracy of Eq.\ \eqref{alphaeff}, we can in fact go beyond the leading order in $\alpha$ in the determination of the edge singularity exponents for both $A(\nu), \chi(\nu)$:  a generic diagram at $n$-th order in $\alpha$ will incur a factor
\begin{align} 
\alpha^n \prod_{l} \text{Tr}\{(P^c)^{n_l}\} \ , \quad n = \sum_l n_l,  \quad n_l  > 1  \ ,
\end{align}  
 where $l$ is the number of $c$-fermion loops, and the condition $n_l > 1$ excludes tadpole-type diagrams which can shift the respective thresholds only. To the leading order in $x_c$, we have 
 \begin{align} 
 \text{Tr}\{ (P^c)^{n_l}\} = \sum_{i} (\lambda^i)^{n_l} \simeq  1 - n_l \int_{\k,\p} \frac{|| \k - \p ||_c^2}{2} , 
  \end{align}
 where we used Eq.\ \eqref{lambdamaxresult}, and the fact that $\lambda_i = O(x_c)$ for $\lambda_i < \lambda_\text{max}$ (App.\ \ref{eigenvalueshiftapp}). As a result, 
 \begin{align} \notag 
\alpha^n \prod_{l} \text{Tr}\{(P^c)^{n_l}\} &\simeq  \alpha^n\left(1 - n \int_{\k,\p} \frac{|| \k - \p ||_c^2}{2}\right) \\  &\simeq \alpha_\text{eff}^n + O(x_c^2) .  
\end{align}
Since $\alpha_\text{eff}$ enters \textit{every} diagram, and not just the logarithmically dominant parquet diagrams, it will determine the Fermi-level scattering phase shift of the $c$-fermions. Therefore we can apply the results expressed in terms of the $s$-wave phase shift, Eq.\ \eqref{strongcouplingps}, with $\delta = \delta(\alpha_\text{eff}) + O(x_c^2)$.

\section{Full band geometry}
\label{fullgeom}

So far, we have studied a structureless $f$-hole. To make the connection to  correlated flat band materials, we must lift this restriction, and reintroduce the full interaction $H_\text{int}$ from Eq.\ \eqref{fullsimpleinteraction}. 

This step is more involved than the introduction of $c$-band geometry alone: if the $f$-hole has a momentum-dependent Bloch function, the problem looses its single-particle character. {Therefore, an analysis of a time-dependent scattering problem as in App.\ \ref{Sapp} does not go through in general}.  However, as long as the band geometry of both $c$- and $f$-bands  is weak, large logarithms remain, and we can still gain insight on spectra by diagrammatic analysis. To keep the problem under control, we will  work to leading order  $O(x) \equiv O(x_c, x_f)$, where $x_f$ quantifies the band geometry of the $f$-band $x_f$  analogously to $x_c$.

\begin{figure}
\centering
\includegraphics[width= \columnwidth]{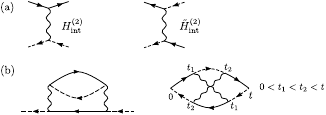}
\caption{(a) Interaction processes that contribute in the limit $E_0 \rightarrow \infty$. (b) Diagrams that involve $\tilde{H}_\text{int}^{(2)}$. For the diagram on the right, the time-domain structure is indicated. Note that interactions are instantaneous.}
\label{newdiags}
\end{figure}

In general, the interaction  $H_\text{int}$ contains terms with up to 4 $f$-operators. In the limit $E_0 \rightarrow \infty$ only terms which conserve the number of $f$-electrons survive: Processes violating this condition are strongly off-shell and suppressed by factors of $1/E_0$. This leaves terms with $4, 0$ or $2$ $f$-operators. As before, terms with $0$ $f$-operators can be absorbed in a renormalized $c$-band, while terms with $4$ $f$-operators are ineffective for a filled $f$-band. We are left with the terms involving two $f$- and two $c$-fermions  graphically represented in Fig.\ \ref{newdiags}.

In addition to the conventional term $H_\text{int}^{(2)}$, processes $\tilde{H}_\text{int}^{(2)}$ where $f$- and $c$-electrons interconvert are allowed, see Fig.\ \ref{newdiags}(a). These processes can appear in the diagrams for $F,\Pi$, as shown in Fig.\ \ref{newdiags}(b): when a conventional diagram contains a $c$-band hole propagating backwards in time, we can replace it with an $f$-band hole. On the right hand side of  Fig.\ \ref{newdiags}(b), we illustrate the time-domain structure of the diagram: As required, all $f$-holes propagate backwards in time. 
While such  processes survive the limit $E_0 \rightarrow \infty$, they are small for a weak band geometry: one can easily  show (see Appendix \ref{upperboundapp}) that for $\k, \p = O(k_F)$, the squared overlap element between $c$ and $f$ bands fulfills $|\!\braket{c,\k | f, \p}\!|^2 = O(x)$. 
Each diagram involving $\tilde{H}_\text{int}^{(2)}$ contains at least four overlap elements, and is therefore of order $O(x^2)$, which is to be neglected within our approximation. Therefore, we only need to keep an interaction term 
\begin{align} 
\label{intterms1}
&H_\text{int} \simeq H_\text{int}^{(2)} =    \\ & \notag  \frac{1}{\Omega} \sum_{ \k, \k^\prime, \q } V_0 c^\dagger_{\k}  f^\dagger_{\k^\prime}  f_{\k^\prime - \q}  c_{\k + \q}   \braket{c,\k| c,\k + \q} \braket{ f,\k^\prime | f,  \k^\prime - \q} \ . 
\end{align}

\subsection{Single-hole spectrum $A$}

\begin{figure}
\centering
\includegraphics[width= 0.6\columnwidth]{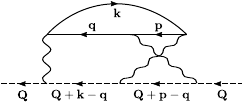}
\caption{Third order diagram for the hole-propagator with  momentum labels.}
\label{genericdiag}
\end{figure}

Via photoemission, a single hole with external momentum $\Q$ can be created. To compute the spectrum $A_\Q$ as defined in Eq.\ \eqref{singleholespec} to order $O(x)$, it is instructive to examine the generic third order diagram shown in Fig.\ \ref{genericdiag}. Given the interaction $H_\text{int}^{(2)}$, the Bloch-overlap structure of this diagram can be written as 
\begin{align} 
\label{BO}
\text{BO}(\Q, \k, \p, \q) \equiv \text{Tr} \{ P^c_\k P^c_\p P^c_\q \}  \text{Tr} \{ P^f_\Q P^f_{\Q + \k - \q} P^f_{\Q + \p - \q} \} , 
\end{align} 
where $P^f_\Q = \ket{f,\Q}\!\bra{f,\Q}$. We can distinguish two cases: First, $\Q = O(k_F)$. In this case, we can expand the projectors $P^f$ in the second factor in small momenta around $\bn$ (again, small in the sense that $x_f \ll 1$). One can then easily show (App.\ \ref{Qdropout}) that the $\Q$-dependence drops out: at $O(x_f)$,  the overlap function depends on the transferred momenta only. This implies that, at order $O(x_f)$, \textit{no dispersion is generated} for the $f$-band on the relevant recoil momentum scale $k_F$. As a result, the large logarithms in the problem are not cut off by the hole recoil, and the power laws in the spectrum remain; as in the previous section, what is left to do is to determine the prefactors of the dominant diagrams by evaluating the overlaps on the Fermi surface.

We may also consider the case $|\Q| \gg k_F$. In this situation, we cannot expand $P^f_\Q$ around $\Q = 0$ in general. Instead, we can expand all $f$-propagators around the momentum $\Q$; the only difference to the preceding case is that the resulting overlap factors  depend on $g^f(\Q)$ instead of $g^f(\bn)$. 

To evaluate the overlap factors involving the $P^f$-projectors, for simplicity we consider an inversion-symmetric Fermi surface (as in the Lieb lattice case), for which
\begin{align} 
\label{nomixed}
g_{ij}^f \int_{\k,\p} k_i p_j = 0 \ . 
\end{align} 
and therefore 
\begin{align} \notag 
g_{ij}^f(\Q) \int_{\k,\p} (k - p)_i (k - p)_j  &=   2 g^f_{ij}(\Q) \int_{\k} k_i k_j  \\ &\equiv 2 \int_k || \k ||^2_{f,\Q}  \ . 
\end{align} 

For the Fermi-surface averaged Bloch overlap of the third-order diagram from Eq.\ \eqref{BO} we then obtain (see App.\ \ref{effintapp})
\begin{align} 
\int_{\k,\p,\q} \hspace{-1em} \text{BO}(\Q, \k, \p, \q) = 1 - 3 \int_{\k} \left( || \k ||^2_c + || \k ||^2_{f,\Q} \right) + O(x^2)  . 
\end{align} 
Likewise, at $n$-th order we obtain a factor 
\begin{align} 
\label{correctionpileup}
1 - n \int_{\k} \left( || \k ||^2_c + || \k ||^2_{f,\Q} \right) \ . 
\end{align} 
This shows that, similar to the case of a trivial $f$-band, for a weak band geometry we can introduce an effective interaction constant
\begin{align} 
\label{alphaeffQ}
\alpha_\text{eff}(\Q) \equiv \alpha \left(1 - \int_{\k}  || \k ||^2_c + || \k ||^2_{f,\Q}  \right) ,  
\end{align} 
which again characterizes an effective momentum-independent scattering problem with phase shift $\delta[\alpha_\text{eff}(\Q)]$. {Although we do not show it here, this suggests that at order $O(x)$  a mapping of the  interaction $H_\text{int}^{(2)}$ to a scattering problem might be possible directly on the level of the Hamiltonian. 

The spectrum resulting from the phase shift $\delta[\alpha_\text{eff}(\Q)]$ takes the form } 
\begin{align} 
A(\Q, \nu) \sim \nu^{2(\delta[\alpha_\text{eff}(\Q)]/\pi)^2 -1} \simeq \nu^{2\alpha_\text{eff}(\Q)^2 -1} \quad \text{for} \  \alpha_\text{eff} \ll 1 . 
\end{align} 

\begin{figure}
\centering
\includegraphics[width= \columnwidth]{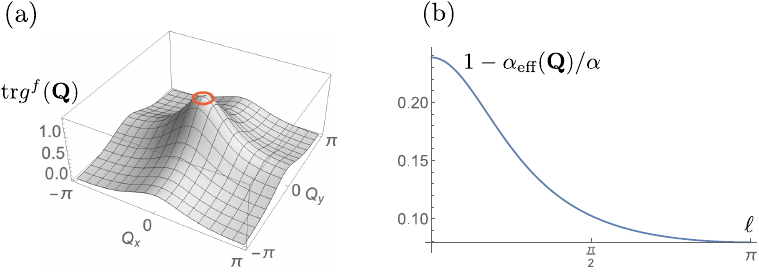}
\caption{(a) $\text{tr} g^f(\Q) = 2\text{tr} g^c(\Q)$ for the Lieb lattice at $t^\prime = 0.6$. The red circle shows the Fermi surface used for Fig.\ (b). Momentum-dependent correction to the effective interaction $\alpha_\text{eff}$, see Eq.\ \eqref{alphaeffQ}, when $\Q$ is varied along the unit cell diagonal $\Q = (\ell,\ell)$,  for  $\mu =  0.015$.   }
\label{Liebnum3}
\end{figure}

The scattering of $f$-fermions to different momentum states on the Fermi surface further reduces the effective interaction $\alpha_\text{eff}$, parametrized by the  distance of the scattered $f$-states in Hilbert space. The photoemission power law exponent inherits the $\Q$-dependence of the local $f$-metric. Results for the Lieb lattice are shown in Fig.\ \ref{Liebnum3}. Note again that, for $\alpha_\text{eff}(\Q)$ to be valid, $x_c,x_f \ll 1$ are required. A closely related requirement is that the metric does not change too strongly on the scale of $k_F$. For the Lieb lattice, this condition breaks down when $t^\prime$ becomes too small and $\text{tr} g^f(\Q)$ is strongly peaked at the $M$-point.

\begin{figure}
\centering
\includegraphics[width= .7\columnwidth]{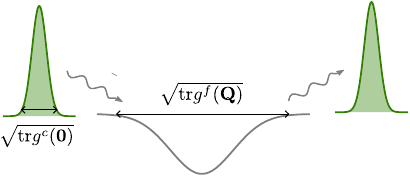}
\caption{Flat band edge singularity as effective potential scattering problem: a $c$ wave packet of width  $\sqrt{\text{tr}g^c(\bn)}$ is scattered by a potential created by an $f$ electron of with effective range $\sqrt{\text{tr}g^f(\Q)}$.  }
\label{scatteringcartoon}
\end{figure}

In a typical potential scattering problem, the phase shift is decreasing when potential range is increased for fixed  potential depth. Recalling the connection of the metric and the minimal spread of the Wannier functions, we can therefore view Eq.\ \eqref{alphaeffQ} as a result of two effects: first, the effective $f$ potential seen by the $c$ fermions has a finite range $\sim \sqrt{ \text{tr} g^f(\Q)}$. Second, any wave packed formed by the $c$ fermions has a finite width $\sim \sqrt{ \text{tr} g^c}$, which effectively adds to the potential range. A cartoon of this effect is shown in Fig.\ \ref{scatteringcartoon}. 

Note that, strictly speaking, it is the Brillouin-zone average of the metric and not the metric itself which determines the spread of the Wannier functions; the simple picture above holds if the momentum scale over which $\text{tr}g$  changes appreciably is much larger than $\sqrt{\text{tr} g}$ itself, which is realized ideally for a system with a uniform metric, e.g., a model with wavefunctions as in a Lowest Landau Level \cite{PhysRevLett.127.246403}.

\subsection{Particle-hole spectrum $\chi$}

For the particle-hole spectrum $\chi(\nu)$, at order $O(x)$ we can follow the same approach as in the previous section and evaluate the Bloch-overlaps for the logarithmic diagrams of Fig.\ \ref{Diagfigfirst}(b), including the interband current matrix elements, see Eq.\ \eqref{interbandcurr}. It is convenient to introduce an operator $\hat{J}^{\eta}_{cf}(\k) = P_\k^c \partial_\eta P^f_\k$. Then, for instance the ``crossed" diagram $\Pi_b^{(2)}$  incures a factor 
\begin{align} 
\label{chioverlap}
\int_{\k,\p,\q} \text{Tr}\{ \hat{J}^\eta_{cf} (\k) P_f(\q + \k - \p) \hat{J}^\eta_{fc}(\q) P_c(\q) \} \ . 
\end{align} 
A challenge in evaluating Eq.\ \eqref{chioverlap} is the momentum-dependence of the current operators. If we set $\hat{J}^\eta(\k) \simeq \hat{J}^\eta(\bn) + \Delta \hat{J}^\eta(\k)$, and neglect the momentum-dependent correction ${\Delta \hat{J}^\eta(\k)}$, the $O(x)$ evaluation of \eqref{chioverlap} and subsequent diagrams, including multiloop processes, proceeds as for the single-hole spectrum, and we find 
\begin{align} 
\label{chifullresult}
\chi(\nu) \simeq \nu^{ -2\delta[\alpha_\text{eff}(\bn)]/\pi + 2(\delta[\alpha_\text{eff}(\bn)]/\pi)^2} \simeq \nu^{ -2\alpha_\text{eff}(\bn)}, 
\end{align} 
with $\alpha_\text{eff}(\bn)$ as in Eq.\ \eqref{alphaeffQ}. A priori, the missed correction $\sim \int_\k \braket{\bn|\Delta \hat{J}^\eta(\k)|\bn}$ is not smaller than $O(x)$. However, this correction only appears twice in each diagram, independent of the order in the interaction, while the corrections derived from the Bloch overlaps for the momenta pile up as in Eq.\ \eqref{correctionpileup}. As a result, the \textit{exponent} of the edge singularity in Eq.\ \eqref{chifullresult} is not impacted by the momentum-dependence of $\hat{J}^\eta$.

\section{Discussion}
\label{discsec}

While the absorption $\chi\sim \nu^{-2\alpha_\text{eff}(\bn)}$ in principle is of direct relevance for correlated flat band materials (e.g.  Moir\'e systems) with a filled flat and doped conduction band, experimental observation might be challenging due to the insufficient energy resolution of spectral measurements to date. Furthermore, in such systems   processes where $c$ electrons relax into the $f$ band must be considered, which lead to an IR tail in the $f$ spectra. Due to energy conservation, these processes  require emission of an excited particle, for instance a photon, phonon or additional $c$ electron (Auger process); they are often suppressed by small matrix elements if the band gap between the $f$ and $c$ band is sufficiently large. 

The case for the observation of the single-hole edge singularity $A(\nu)$ can be made more easily, in particular when $c$-particles and $f$-hole  correspond to different particle species and thus cannot interconvert, which prohibits the relaxation process described above. This is possible if  the $f$-hole is an ``impurity" coupled to a $c$-fermion bath, a scenario that can, e.g., be realized in experiments involving quantum dots, where edge singularities have  been observed for trivial bands \cite{JanQuantumQuench}. Likewise, this situation applies to typical ``polaronic" spectral measurements in ultracold gases, where two different particle species are studied (see e.g.\  \cite{PhysRevLett.99.170404, PhysRevA.102.023304,schmidt2018universal} and Refs.\ therein).  In this case, a mature experimental technique for measuring single-particle spectra is inverse (injection) RF spectroscopy \footnote{standard RF spectroscopy, which is akin to ARPES, measures the occupied  part of the spectral function, which complicates the detection of impurity spectra}: here, the $f$-particles have to be initially prepared in a state where the interaction with the $c$ particles can be neglected; by applying a weak RF pulse of frequency $\omega$ ($\nu$ plus threshold energy), particles are excited into an interacting $f$-band. In linear response, the depletion current from the non-interacting state is proportional to $A(\nu)$, and the spectral resolution is inversely proportional to the pulse duration. A possible observable is the dependence of the spectrum on the Fermi surface volume, either via Eq.\ \eqref{Anucgeom} for general Fermi momentum, or Eq.\ \eqref{Anunorm} for small one (separating out the distinct $k_F$-dependence of $\alpha$). For the ultracold gases, a crucial experimental challenge is the required small temperature $T \ll \mu$; finite temperatures will broaden the edge singularities in a well-understood manner \cite{ PhysRevB.73.075122, PhysRevX.2.041020}.

If the $f$-hole is part of a flat band with non-vanishing band geometry, the momentum dependence of $A_\Q$ can in principle be extracted by combining the RF measurement  techniques such as time-of-flight mapping \cite{PhysRevLett.109.095302} or Raman spectroscopy \cite{PhysRevX.10.041019}. If the power law exponent $\alpha_\text{eff}(\Q)$ can be extracted from such a measurement, the flat band metric $g^f(\Q)$ can be mapped out in experiment. When such momentum resolution is not available, the RF spectrum yields   the  Brillouin zone average of the $f$-spectrum; if the non-interacting $f$-state has flat dispersion as well, simply $A_{\text{RF}}(\nu) \sim \int d\Q A_\Q(\nu)$. At $O(x)$, we can equivalently average the momentum-dependent exponent $\alpha_\text{eff}(\Q)$. 
Therefore, at weak doping the momentum-averaged RF measurement of the power law exponent gives access to the momentum-averaged metric, a  probe of the minimal real-space spread of the associated Wannier functions. 

The universal effective interaction $\alpha_\text{eff}(\Q)$ was obtained for weak doping. At strong doping, the universality breaks down, and the results will strongly depend on lattice details. In particular, if the scale on which $\text{tr} g^f(\Q)$ changes strongly becomes  $O(k_F)$,  effective mass generation becomes important, cutting off logarithmic singularities. In passing, we note that effective two-body masses in the flat band can be related to the integral over the quantum metric (which can be non-zero even for a momentum-independent metric) if a {sublattice-sensitive} contact interaction is used \cite{PhysRevB.98.220511, PeottaSuperfluidity, PhysRevA.103.053311}. For a finite $f$-band mass, the $f$-particle becomes a mobile ``Fermi polaron", which can for instance be described by variational methods \cite{PhysRevA.74.063628} and has  been explored in the literature in various lattice contexts \cite{PhysRevB.94.235105, Grusdtoppol2016, PhysRevB.99.081105, PhysRevB.100.075126, PhysRevA.102.033305, PhysRevB.103.245106}.

\section{Summary and Outlook}
\label{concsec}

In this work, we derived a universal lattice generalization of edge singularities: we considered processes where a single degree of freedom (hole) in a $f$-band interacts with fermions in a dispersive $c$-band, allowing for a non-trivial Bloch geometry for both bands, and evaluated corresponding single-hole  $\sim \braket{f^\dagger f}$ and particle-hole $\sim \braket{c f^\dagger f c^\dagger}$ spectra.   We found that the leading effect of the band geometry is to reduce the effective coupling that enters the edge singularity exponents, which we derived by evaluating Bloch-function overlaps appearing in the respective singular Feynman diagrams. For $k_F$ much smaller than a reciprocal lattice vector, corrections to the exponents are proportional to the quantum metrics times the Fermi energy, which can be traced back to the finite range of the effective scattering potential created by the $f$-hole. Our results for the exponents are summarized in Tables \ref{tableA}, \ref{tablechi}.

While we considered a two-band scenario, real materials can feature degenerate $c$- or $f$-bands. We expect our results to carry over to this situation as well, replacing single-band projectors by projectors on a degenerate set of bands, and the (abelian) quantum metric by the non-abelian one \cite{HerzogArbeitman2024,PhysRevB.81.245129}

In our derivation of the spectra, diagrammatic perturbation theory and a coordinate representation of the metric was employed. {For a trivial impurity but non-trivial $c$-band, we have cross-validated these results by perturbatively evaluating the $S$-matrix of the associated scattering problem. One could also attempt a full evaluation of the $S$-matrix  by solving a Lippmann-Schwinger equation. Furthermore, it} would be interesting to reformulate the problem with a trivial impurity but non-trivial $c$-band purely as a scattering problem on Riemannian manifold which reflects the non-trival Bloch band geometry \cite{Riemannresp}; this formulation may also enable a numerically exact solution valid for all frequencies $\nu > 0$ via {bosonization and Functional Determinant methods} \cite{PhysRevB.81.085436, Gutman_2011, schmidt2018universal}. 

Lastly, we note that the orthogonality catastrophe underlying the edge singularities can occur in bosonic systems as well \cite{PhysRevA.103.013317}, which is of particular relevance for the ultra-cold gas setups. To enrich this bosonic orthogonality with effects of lattice geometry and topology is a worthwhile goal for future study. 

\textit{Acknowledgement}.  
I thank Moshe Goldstein, Jonah Herzog-Arbeitman,  Dan Mao and Erich J. Mueller for useful discussions, and am especially grateful to Debanjan Chowdhury for insightful suggestions and support during the preparation of this manuscript. {Furthermore, I thank an anonymous referee for several useful literature suggestions}. I acknowledge funding by the German Research Foundation (DFG) under Project-ID 442134789.

\renewcommand{\arraystretch}{1.5}
\begin{table*}
\centering
\begin{tabular}{|c |c |c|}
\hline
 $\boldsymbol{A(\nu)}  $ \ & Exact in interaction  \  & \ $O(\alpha^2) $ \   \\[0.3em] 
\hline 
No band geometry\ & $2[\delta(\alpha) /\pi]^2 -1 $ &  $2\alpha^2 - 1$ 
\\ \hline
$c$-band geometry &  $2[\delta(\alpha_\text{eff}) /\pi]^2 -1 $ &  $2\alpha^2\text{Tr}\{(P_c)^2\}- 1$ \\
& $
\alpha_\text{eff} \simeq \alpha\left(1 - \text{tr}g^c(\bn) \frac{k_F^2}{d} \right)$ & $P_c = \int_\k \ket{c,\k} \!\bra{c,\k}$ \\ \hline
$c$- and $f$-band geometry &  $2[\delta(\alpha_\text{eff}(\Q))/\pi]^2 -1$& $2\alpha^2_\text{eff}(\Q)  -1$ \\
& $
\alpha_\text{eff}(\Q) \simeq \alpha\left(1 - \text{tr}g^c(\bn) \frac{k_F^2}{d}  -  \text{tr} g^f(\Q) \frac{k_F^2}{d} \right)$ &  \\
 [1ex]
\hline
\end{tabular}
\caption{Exponents $\gamma$ for the single-hole spectrum $A(\nu) \sim \text{Im}\left[ \braket { f^\dagger f } (\nu) \right] \sim \nu^\gamma$ in $d$ dimensions, relevant for photoemission. Results involving the metric are for a spherical Fermi surface. }
\label{tableA}
\end{table*}

\begin{table*}
\centering
\begin{tabular}{|c |c |c|}
\hline
 $\boldsymbol{\chi(\nu)}$ \ & Exact in interaction  \  & \ $O(\alpha) $ \   \\[0.3em] 
\hline 
No band geometry\ & $-2\delta(\alpha) /\pi + 2[\delta(\alpha)/\pi]^2$ &  $-2\alpha$ 
\\ \hline
$c$-band geometry &  $-2\delta(\alpha_\text{eff}) /\pi + 2[\delta(\alpha_\text{eff})/\pi]^2$ &  $- 2\alpha \lambda_\text{max}$ \\
&  & $\lambda_\text{max} = \text{max}[\text{EV}(P_c)]$ \\ \hline
$c$- and $f$-band geometry &  $ -2\delta(\alpha_\text{eff}(\bn)) /\pi + 2[\delta(\alpha_\text{eff}(\bn))/\pi]^2
$ & $-2\alpha_\text{eff}(\bn) $  \\
 [1ex]
\hline
\end{tabular}
\caption{Exponents $\gamma$ for the particle-hole spectrum $\chi(\nu) \sim  \text{Im}\left[ \braket { c f^\dagger f c^\dagger } (\nu) \right] \sim \nu^\gamma$ in $d$ dimensions, relevant for interband absorption. For definitions of $\alpha_\text{eff}, P_c$, see Table \ref{tableA}.}
\label{tablechi}
\end{table*}

\appendix 
\section{Eigenvalue shift}
\label{eigenvalueshiftapp}

To determine the maximal Eigenvalue $\lambda_\text{max}$ of the Fermi-surface averaged projector $P^c$, we perform an expansion
\begin{align} 
\label{Tnotation}
P^c &= \int_\k \ \ket{\k}\!\bra{\k} \simeq \ket{\bn}\!\bra{\bn} + T_1 + T_2 \\  \notag
T_1 &=  \int_\k k_i \left( \ket{i}\!\bra{\bn} + \ket{\bn}\!\bra{i} \right)   \\   
T_2 &=  \int_{\k} \frac{k_i k_j}{2} \left( \ket{ij}\!\bra{\bn} + \ket{\bn}\!\bra{ij} + \ket{i}\!\bra{j} + \ket{j}\!\bra{i}\right) \notag \ ,  
\end{align} 
where we suppressed the label $c$ in the Bloch functions. Retaining terms up to second order in the expansion of the Bloch functions,  the maximal eigenvalue reads 
\begin{align} 
\lambda_\text{max} \simeq 1 + \braket{\bn| T_1 + T_2 |\bn} + \sum_{\gamma} {|\braket{\gamma | T_1 | \bn }|^2}, 
\end{align} 
where $\ket{\gamma}$ are vectors orthogonal to $\ket{ \bn}$. We have $\braket{\bn | T_1 | \bn} = 0$ and  
\begin{align} 
\braket{ \bn | T_2 | \bn } = - \int_\k g_{ij}^c k_i k_j. 
\end{align} 
Furthermore, 
\begin{align} \notag
&\sum_{\gamma} |\braket{\gamma | T_1 | \bn }|^2  = \braket{ \bn | T_1 | \left( \mathbbm{1} - \ket{\bn}\!\bra{\bn}  \right) T_1 | \bn}   = \braket{\bn | T_1 T_1 |\bn}   \\ \notag &= \int_{\k,\p} k_i p_j \left( \braket{\bn|i}\braket{\bn|j} + \braket{\bn|i} \braket{j|\bn} + \braket{i|j} + \braket{i|\bn} \braket{j|\bn} \right) \\ &=  g_{ij}^c \int_{\k,\p} k_i p_j , \label{T1squared}
\end{align} 
where in the last step the first two terms cancel and the last two terms can be symmetrized by relabeling $\k \leftrightarrow \p$. Therefore 
\begin{align} \notag
\lambda_\text{max} &\simeq 1 -\frac{1}{2} g_{ij}^{c} \int_{\k,\p} (k - p)_i (k - p)_j \\  &\equiv 1 - \int_{\k,\p} \frac{|| \k - \p ||_c^2}{2} , \label{lambdashiftfinal}
\end{align} 
as in Eq.\ \eqref{lambdamaxresult} of the main text. 

This result also allows to estimate the remaining Eigenvalues of $P^c$ which appear in the general expression for $\chi(\nu)$, Eq.\ \eqref{chinucgeom}. We note that  
\begin{align} 
\text{Tr}\{ P^c \}= \int_\k \text{Tr} \{ \ket{\k}\! \bra{\k} \} = \int_\k = 1 =  \lambda_\text{max} + \sum_{\lambda_i < \lambda_\text{max}} \lambda_i \ . 
\end{align}
Therefore
\begin{align} 
\label{otherEV}
\sum_{\lambda_i< \lambda_\text{max}} \lambda_i  \simeq  \int_{\k,\p} \frac{|| \k - \p ||_c^2}{2} 
\end{align} 
Note also that $P^c$ is positive semidefinite, therefore $\lambda_i > 0  \ \forall i$. 

\section{Short Proofs}
\label{proofapp}

\subsection{Upper bound for $|\! \braket{c,\k | f , \p} \!|^2$ }
\label{upperboundapp}
We have 
\begin{align} 
&1 = \braket{c, \k| c, \k } = \bra{c,\k} \left(\sum_n \ket{n,\p}\! \bra{n,\p}\right) \ket{c,\k} \geq  \notag  \\ \notag 
&  |\!\braket{c,\k | f,\p}\!|^2 +  |\!\braket{c,\k | c,\p}\!|^2 =  |\!\braket{c,\k | f,\p}\!|^2 + 1 - O(x_c) , 
\end{align} 
therefore 
\begin{align} 
 |\!\braket{c,\k | f,\p}\!|^2 = O(x_c) \ .
 \end{align} 
 In the same manner, one can also show that 
\begin{align} 
 |\!\braket{c,\k | f,\p}\!|^2 = O(x_f), \quad x^f = k_F^2 \text{tr} g^f(\bn)  \  . 
 \end{align} 
 
 \subsection{$\Q$-independence of overlap factors for $\Q = O(k_F)$}
 \label{Qdropout}
 
 Consider the Bloch-overlap factor $\text{BO}_f(\Q) \equiv \text{Tr} \{ P^f_\Q P^f_{\Q + \k - \q} P^f_{\Q + \p - \q} \}$ from Eq.\ \eqref{BO}. To shorten notation, we write $\a = \k - \q$, $\b = \p - \q$. When $\Q = O(k_F)$, we can expand all projectors to second order in momenta with notation similar to Eq.\ \eqref{Tnotation}: 
 \begin{align}
 \label{Pfexpansion}
  &P^f_\Q \simeq  \ket{f, \bn}\! \bra{f, \bn} + Q_i T_1^i + Q_i Q_j T_2^{ij},  \\  \notag
  &T_1^i =  \left( \ket{f, i}\!\bra{f, \bn} + \ket{f, \bn}\!\bra{f, i} \right)  \\ \notag 
  &T_2^{ij} = \\   &\frac{1}{2} \left( \ket{f,ij}\!\bra{f,\bn} + \ket{f,\bn}\!\bra{f,ij} + \ket{f,i}\!\bra{f,j} + \ket{f,j}\!\bra{f,i} \right) \notag 
   \end{align} 
 Collecting all second-order terms (first order terms vanish), we obtain 
 \begin{align} 
& \text{BO}_f(\Q) \simeq  \braket{f, \bn| T_1^i T_1^j | f, \bn} \times  \\ \notag &\left( Q_i (Q + a)_j + (Q + a)_i (Q + b)_j + (Q + b)_i Q_j \right) +   \\ \notag &\braket{f, \bn|T_2^{ij}  | f, \bn}  \times \\ \notag &\left( Q_i Q_j + (Q + a)_i (Q + a)_j + (Q + b)_i (Q + b)_j \right) \ . 
 \end{align} 
Using that $\braket{f, \bn|  \frac{1}{2} \left(T_1^i T_1^j + T_1^j T_1^i  \right) | f, \bn} = g_{ij}^f$, see Eq.\ \eqref{T1squared}, and $\braket{f, \bn|T_2^{ij}  | f, \bn} = - g_{ij}^f$, we immediately see that all terms involving $Q$ appear in symmetrical combinations and therefore cancel. From the  form of $\text{BO}_f(\Q)$ it is clear that this property will also generalize to higher order. 
 
 \section{Effective interaction for full band geometry} 
 \label{effintapp}
 
 To illustrate the derivation of 
\begin{align} 
\label{alphaefffullapp}
\alpha_\text{eff} \equiv \alpha \left(1 - \int_{\k}  || \k ||^2_c + || \k ||^2_{f,\Q}  \right), 
\end{align} 
we consider the third order Bloch overlap 
\begin{align}
\label{BOeval}
\text{BO}(\Q) \equiv \int_{\k,\p,\q} \text{Tr} \{ P^c_\k P^c_\p P^c_\q \} \times \text{Tr} \{ P^f_\Q P^f_{\Q + \k - \q} P^f_{\Q + \p - \q} \} . 
\end{align} 
We evaluate this expression by expanding the projectors up to second order in momenta as in Eq.\ \eqref{Pfexpansion}.  Under the assumption of an inversion-symmetric Fermi surface, where mixed terms in momenta vanish [Eq.\ \eqref{nomixed}], this expansion can be performed by sequentially keeping one of the momenta $\k, \p, \q$ non-zero, and setting the remaining ones to zero; furthermore, we only need to keep the terms of the form $T_2^{ij}$ from Eq.\ \eqref{Pfexpansion}. Lastly, at order $O(x)$, the corrections from the two traces in Eq.\ \eqref{BOeval} add up and can be evaluated independently. 

\begin{figure}
\centering
\includegraphics[width= 0.75\columnwidth]{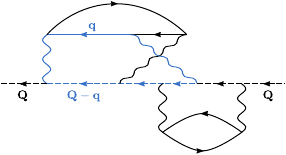}
\caption{Generic diagram with one $c$-fermion momentum $\q$ set to non-zero only. The momentum $\q$ runs in a a loop through the diagram -- the propagators and interaction lines which transport $\q$ are marked blue. This ensures that all projectors $P^f_{\Q - \q}$ in the trace are adjacent to each other. }
\label{nonzeromom}
\end{figure}

As a result, we obtain 
\begin{align} 
\int_{\k,\p,\q} \text{Tr}\{P_\k^c P_\p^c P_\q^c\} = 1 - 3 \int_\k ||\k||_c^2 + O((x_c)^2)
\end{align} 
For the $f$-projector trace, a possible difference is that the momentum $-\q$ appears in two projectors; however, since these projectors are adjacent, upon setting $\k = \p = 0$ as discussed above, we have $P^f_{\Q -\q} P^f_{\Q - \q} = P^f_{\Q - \q}$, and the $\q$-integral therefore gives the same contribution as the $\k,\p$-integrals. Thus 
 \begin{align} 
\text{BO}(\Q) = 1 -3\int_{\k} \left( || \k ||^2_c + || \k ||^2_{f,\Q} \right) \ . 
\end{align} 
For a generic $n$-th order diagram (excluding tadpole-diagrams), the evaluation proceeds in the same manner, including diagram which contain multiple $c$-fermion loops.  In particular, one can easily convince oneself that ``non-zero" momenta only appear in adjacent $f$-projectors; see Fig.\ \ref{nonzeromom} for an illustration of this fact.

 Therefore, the corresponding overlap reads, 
 \begin{align} 
 \text{BO}_n(\Q) = 1 - n \int_\k \left( || \k ||^2_c + || \k ||^2_{f,\Q} \right) \ .
 \end{align} 
Including the coupling constant $\alpha$, we have, at $n$-th order 
 \begin{align}
 \alpha^n \left[ 1- n \int_\k \left( || \k ||^2_c + || \k ||^2_{f,\Q} \right) \right] = \alpha_\text{eff}^n + O(x^2) , 
 \end{align} 
 where $\alpha_\text{eff}$ is given in Eq.\ \eqref{alphaefffullapp}.

 {\section{$S$-matrix approach for trivial $f$-band}
 \label{Sapp}
 As shown in Refs.\ \cite{PhysRevLett.91.266602, PhysRevB.71.045326}, scattering problems with a time-dependent potential can be solved by mapping to a Riemann-Hilbert boundary value problem, which also allows for generalizations to non-equilibrium settings \cite{PhysRevLett.91.266602, PhysRevB.71.045326} or finite temperatures \cite{PhysRevB.73.075122}. For the spectra $A(\nu), \chi(\nu)$ of interest to us, such an analysis results in Eqs.  \eqref{partialpsA}, \eqref{partialpschi}: 
 \begin{align}
&A(\nu) \sim \nu^{2\sum_j  (\delta_j/\pi)^2 -1} \label{partialpsAapp} \\
&\chi(\nu) \sim \sum_{j} \nu^{-2\delta_j /\pi + 2 \sum_{j^\prime} (\delta_{j^\prime}/\pi)^2}  \label{partialpschiapp} \ , 
\end{align} 
Here, $\delta_j$ are derived from the Eigenvalues $\exp(2i\delta_j)$ of the $S$-matrix at the Fermi level. 

The $S$-matrix connects in- and outgoing scattering states. Using the potential $H_\text{int}^{(1)}$ from Eq.\ \eqref{singlepartH1}, 
\begin{align} 
H_\text{int}^{(1)} =  - \frac{1}{\Omega} \sum_{\k,\q} V_0 \braket{c,\k + \q|c,\k} \ket{\Psi_{c,\k+\q}}\! \bra{\Psi_{c,\k}} , 
\end{align} 
to the leading order in $V_0$ (Born approximation), its matrix elements can be expanded as 
\begin{align} \notag
S_{\k,\p} &=    \braket{\Psi_\k | \Psi_\p} - 2\pi i \braket{\Psi_\k | H_\text{int}^{(1)} | \Psi_\p } \delta(\epsilon_\k - \epsilon_\p) + O(V_0^2) \\   & =\delta_{\k,\p} - 2\pi i \braket{\k|\p} \delta(\epsilon_\k - \epsilon_\p), 
\end{align} 
where we have used the orthonormality of the full energy Eigenfunctions, and drop $c$-labels for brevity. 

By definition, the $S$-matrix at the Fermi level has Eigenvalues $\exp(2i\delta_j) \simeq 1 + 2i\delta_j$. Therefore, to the leading order in $V_0$, the phase shifts $\delta_j$ can be extracted from the Eigenvalues of the matrix $S^{(1)}$ given by  
\begin{align} 
S^{(1)} = \alpha \int_{\k,\p} \braket{\k|\p} \ket{\Psi_\k} \! \bra{\Psi_\p}, 
\end{align} 
where the $k$-integral is restricted to the Fermi-surface as defined in Eq.\ \eqref{I2}, and $\alpha = \rho V_0$. 

To reproduce the diagrammatic results of Eqs.\  \eqref{Anucgeom} and \eqref{chinucgeom}, we need to show that the Eigenvalues of $S^{(1)}$ and the Eigenvalues of $\alpha P^c = \alpha \int_\k \ket{\k} \! \bra{\k}$ agree. This is not readily obvious:  $P^c$ is a matrix in band (or orbital) space. For the example of the Lieb lattice, it is a $3\times 3$ matrix. On the other hand, nominally $S^{(1)}$ is a matrix in the larger space of energy Eigenstates indexed by momenta on the Fermi surface. However, because the entries of $S^{(1)}$ are determined by the Bloch functions $\ket{\k},\ket{\p}$ which are defined in the smaller band space, the ranks of $S^{(1)}$ and $P^c$ are the same. To see this, we can e.g.\ assume that $\ket{\p}$ is defined in a two-dimensional space and has a decomposition $\ket{\p} = a_p \ket{\a} + b_p \ket{\b}$ into a basis $\{\ket{\a}, \ket{\b}\}$. Assume that we discretize the integral over the Fermi surface into a summation over momenta ${\k_1,\hdots \k_n}$, such that $S^{(1)}$ is an $n\times n$ matrix. Then, all columns of $S^{(1)}$ are spanned by the two vectors $(\braket{\k_1 | \a}, \hdots ,\braket{\k_n |\a})^T$ and $(\braket{\k_1 | \b}, \hdots ,\braket{\k_n |\b})^T$, which shows that $S^{(1)}$ has rank 2. 

To see that the non-zero Eigenvalues of  $S^{(1)}$ and $\alpha P^c$ agree, one can for instance consider traces over matrix powers: 
\begin{align}
&\text{Tr}\{\alpha P^c\} = \alpha \int_\k \text{Tr}\{\ket{\k} \! \bra{\k}\} = \alpha =  \text{Tr}\{S^{(1)}\} \\ 
&\text{Tr}\{\left(\alpha P^c\right)^2\} =  \alpha^2 \int_{\k,\p} \text{Tr}\{\ket{\k} \!\braket{\k|\p} \! \bra{\p}\} = \int_{\k,\p} |\!\braket{\k|\p}\!|^2 \\ \notag
&\text{Tr} \big\{( S^{(1)})^2 \big\} = \\ \notag & \alpha^2 \hspace{-1em} \int\displaylimits_{\q,\k_1, \p_1,\k_2,\p_2} \hspace{-1em}\braket{\Psi_\q | \Psi_{\k_1}} \braket{\Psi_{\p_1} | \Psi_{\k_2} } \braket{\Psi_{\p_2} | \Psi_\q} \braket{\k_1 | \p_1} \braket{ \k_2 | \p_2}   \\ & = \text{Tr}\{\left(\alpha P^c\right)^2\}  \ . 
\end{align} 

Proceeding analogously, one finds that $\forall n$: 
\begin{align} 
\text{Tr}\{(\alpha P^c)^n\} = \sum_i (\alpha \lambda_i)^n =  \text{Tr}\{(S^{(1)})^n\} = \sum_i (\delta_i/\pi)^n
\end{align}
By comparing the largest Eigenvalues of $P^c, S^{(1)}$ (which are real) for large even values of $n$ one can sequentially show that all Eigenvalues are the same. This proves that the power law exponents obtained from  the diagrammatic and $S$-matrix methods agree to the leading order in $\alpha$. 
}

\bibliography{OC_flatband}

\end{document}